# A Virtual Airplane for Fear of Flying Therapy

Larry F. Hodges[α], Barbara O. Rothbaum[β], Benjamin Watson[α],
G. Drew Kessler[α], and Dan Opdyke[γ]

[α]Graphics, Visualization & Usability Center
Georgia Tech

[β]Department of Psychiatry and Behavioral Sciences
Emory University School of Medicine

[γ]Department of Psychology
Georgia State University

## Abstract

*Fear of flying is a serious problem that affects millions of individuals. Exposure therapy for fear of flying is an effective therapy technique. However, exposure therapy is also expensive, logistically difficult to arrange, and presents significant problems of patient confidentiality and potential embarrassment. We have developed a virtual airplane for use in fear of flying therapy. Using the virtual airplane for exposure therapy is a potential solution to many of the current problems of fear of flying exposure therapy. In this paper we describe the design of the virtual airplane and present a case report on its use for fear of flying exposure therapy.*

## Motivation and Background

### Fear of Flying

Fear of flying is a serious problem with growing financial repercussions. It has been estimated that 10-25% of the general population suffer from a fear of flying. A Boeing Airplane Company survey indicated that 25 million adults in the United States are fearful about flying. In addition, approximately 20% of those who do fly depend on alcohol or sedatives during flights. This fear and resulting avoidance of flying has serious consequences–including career repercussions, social embarrassment and restrictions–as well as financial consequences. In 1982, annual revenue loss for the U.S. air travel industry was estimated at 1.6 billion dollars [14].

## Treatment Approaches

Approaches to the treatment of fear of flying include anxiety management methods, provision of accurate information regarding airplanes and flying, and exposure techniques. Actual exposure to flying, usually occurring after the client is provided with initial training in anxiety management methods, has been consistently reported as effective [2, 3, 14, 20]. Exposure is usually provided in stages [7, 8, 14] with clients first practicing going to an airport, seeing and hearing the sights and sounds of airplanes taking off and landing. Subsequently they might actually enter and sit in a stationary airplane. Ideally an actual flight experience would be the capstone of the therapy program.

Given the prevalence and impact of fear of flying, relatively few controlled studies have been conducted. Some fear of flying "programs" exist in large metropolitan cities, often sponsored by airlines and charging high fees to include flights, and usually are not covered by insurance. These programs have not been subjected to rigorous evaluation, and therefore their efficacy cannot be ascertained. Most of those that have been reported in the literature were not conducted with civilian airline passengers, but involved military or airline industry flight crews. The difficulty and expense of using actual airplanes and flights for exposure have daunted many researchers and therapists. Fear of flying therapy involves considerable expense in terms of therapist's time to accompany their client to an airport for exposure, in addition to the cost of buying tickets for actual flights for the client and therapist. Logistically, there are also many obstacles. The therapist might spend considerable time arranging with airline officials to have access to stationary planes. For inflight sessions, patient confidentially and embarrassment are also significant factors.

## Virtual Reality Exposure

Traditional approaches to exposure therapy include imaginal (subject images the stimulus) exposure and *in vivo* (subject is exposed to actual physical situation) exposure. Virtual reality exposure, in which the subject is exposed to a virtual environment containing the feared stimulus, has been shown in a controlled study to be an effective treatment approach for acrophobia–the fear of heights [11, 15, 16].

To extend VR exposure to fear of flying, we designed a virtual airplane which the client experiences by wearing a head-mounted display with stereo earphones. The client receives both visual and auditory cues of actually being on an aircraft. The therapist can see and hear what the subject is experiencing on a TV monitor.

There are several potential advantages in using virtual reality exposure as compared to *in vivo* exposure techniques.

**Therapist Time.** In a major metropolitan area a therapist can easily spend hours commuting to and from the airport in order to conduct a therapy session. Using the virtual airplane for exposure allows therapy sessions to be conducted in the therapists office.

**Control of the Environment.** Most of the elements of exposure therapy for fear of flying are not subject to control by the therapist. A commercial flight is not going to spend an extra hour taxiing on the runway because a passenger is experiencing anxiety about the takeoff. Good weather and a smooth flight can not be guaranteed. Arranging with an airline to make a stationary plane available so the subject can get used to just sitting in the plane on the ground can be difficult or impossible. All of these factors are under direct control of the therapist when using the virtual airplane for exposure.

**Risk.** Although small, the possibility of harm exists with *in vivo* exposure (e.g., driving to the airport, flying on an airplane). Virtual exposure eliminates such risks.

**Patient Confidentiality.** Since therapy is conducted in the therapist's office, virtual airplane exposure allows for complete patient confidentiality and eliminates the possibility of public embarrassment.

**Cost.** Our preliminary virtual airplane was developed in a laboratory environment on a Silicon Graphics Reality Engine, which displayed images on a Virtual Research VR 4 head-mounted display with a 48 degree horizontal field of view. The subject's head position and orientation was tracked with a Polhemous Isotrack 3D tracker. Current advances in personal computer technology indicate that the same performance and capabilities for a virtual airplane will be possible on a PC-based system within a year. The same PC-based system could be used for other types of exposure therapy (such as fear of heights). The decrease in actual therapist time and the elimination of the need for airplane tickets would reduce the cost of therapy to both the therapist and client.

**Willingness to Undergo Therapy.** We have held many discussions with the subjects who have undergone virtual reality therapy in our previous studies and with people who have contacted us seeking virtual reality therapy. A consistent theme has been that many individuals are willing to undergo exposure therapy in virtual environments who would not be willing to undergo therapy by facing the real, physical situations that they fear.

## Design of the Virtual Airplane

Design considerations for the virtual airplane were significantly more complex than those of our previous environments used for virtual reality therapy [11]. Our previous environments–an open elevator, a series of bridges and a series of balconies–were all purely visual, relatively static environments with little or no animation except that provided by the user's head motion. After the design and modeling stage of the environments had been completed, software implementation of the virtual environments required only basic VR software tools to accomplish standard tasks such as body tracking and image generation. The virtual airplane required these capabilities plus significant animation, control, and sound effect capabilities.

The major stages in the development of the virtual airplane were: creation of a visual model of the passenger cabin of a commercial aircraft; creation of an environment for the aircraft to fly through; design of the control sequence software for the different stages of an airplane flight; and design of sound effects.

### Visual Models

To be convincing the airplane model had to be built to the proper general size and scale to match that of a real passenger cabin of a commercial aircraft. We felt that details such as the fabric texture and color of the seats, proper location of seatbelt signs, aisle width, and window locations would be important to providing an appropriate "look and feel" of being in a real aircraft. A commercial airline company allowed us access to their passenger cabin simulators that are used to train flight attendants. A design team consisting of two undergraduate students and a high school teacher who was working with us as a summer intern spent several hours in a full scale model of a Boeing 747 taking photographs of the passenger cabin from all angles and measuring such details as height, width and depth of the seats, ceiling height, aisle width, etc. The photographs and measurements were then used as a guide to create the passenger cabin model using Wavefront modeling software [21]. The interior of the passenger cabin is shown in figure 1.

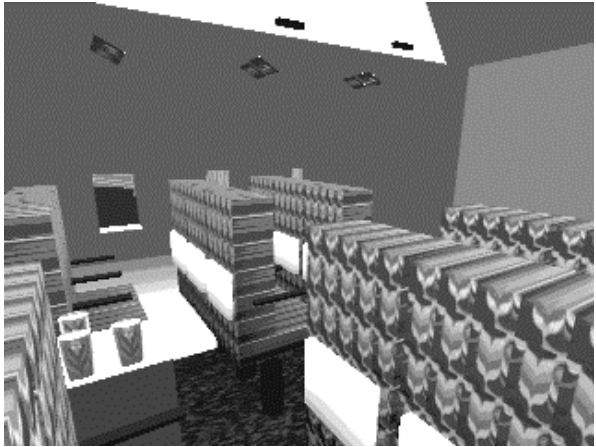

**Figure 1. Passenger cabin of the virtual airplane.**

The second modeling task was to create a world for the aircraft to fly in that could be viewed through a window of the aircraft by the passenger. For the stages when the plane was parked, taxiing, taking off or landing we created a runway and airport model with various types of airport buildings and scenery (figure 2). As the aircraft gains in altitude and distance from the airport, the surface of the earth is represented by texture maps from aerial photographs. Clouds in the sky are also simulated with texture maps. Figure 3 illustrates the passenger's view through the window.

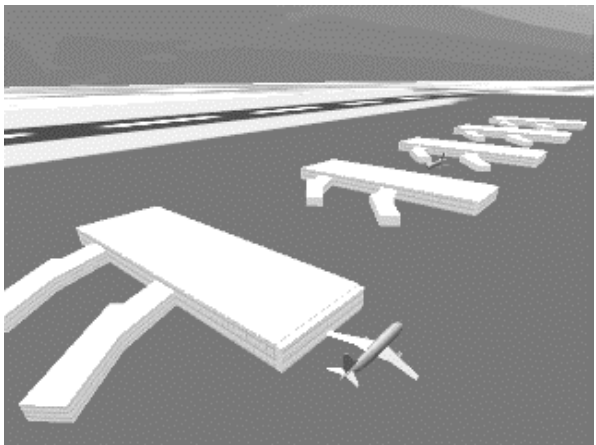

**Figure 2. The airport.**

The path followed by the airplane in each stage of its flight is described by a predefined cubic spline. The location of the airplane on the spline is determined by first finding the time passed since the last plane relocation, relating this to the plane's current speed and thus finding the distance traveled. We then increment the parametric variable by an amount corresponding to this distance, and use the resulting value to sample the spline and find the plane's current location in three dimensions. The orientation of the airplane is controlled by a function related to the speed of the airplane and the first and second derivatives of the spline at the current location.

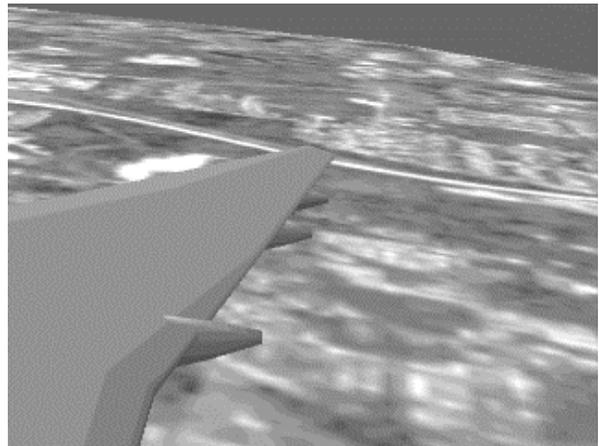

**Figure 3. Passenger's view through the window of the virtual airplane.**

The notion of time used in the simulation is real time. This makes the simulator portable between machines with different processing speeds, and simplifies the synchronization of the simulation graphics and sound effects since sound effects are always played in real time.

**The Flight Sequence**

There is a natural order to events that occurs in a real airplane flight. In general, we expect to sit at the terminal for a while, taxi to a specific spot on the runway, take off, be in flight for an extended time, and land. Some of these events, such as takeoff and landing, are relatively brief events. Other events, such as the actual flight between takeoff and landing, can vary in duration from a few minutes to several hours. Conditions can also vary. The different events can occur in good or bad weather. The passenger's view from the window can vary considerably with time, altitude and weather conditions. Additional variables are introduced by the needs of the therapist and the patient when conducting therapy. An entire therapy session may consist of only taxiing about the runway without ever actually taking off. Several flights in good weather may be required before the patient is ready to fly in a thunderstorm. Our goal was to design software that would simulate the sequence of events of a real flight while also allowing a degree of flexibility to quickly vary conditions and duration of events in response to the needs of the therapy session. For example, if a patient was comfortable taxiing on the runway the therapist could request that the plane takeoff in the next few minutes, or if a patient was no longer nervous while flying in good weather, the therapist could request that the plane fly into a thunderstorm. In either case we wish to move quickly to the next event but do so in a manner that does not disrupt the illusion of being on a real aircraft.

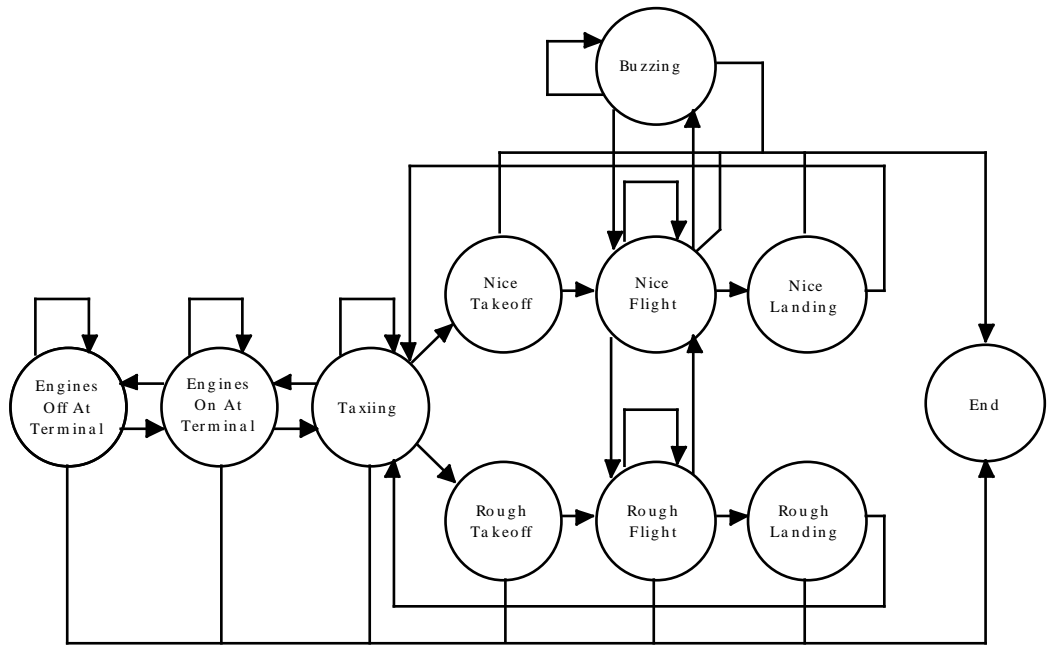

**Figure 4. Airplane State Diagram.** Circles show possible current states of the virtual airplane. Arrows show possible transitions from one state to another.

To balance the need to provide a continuous flight experience while maintaining responsiveness to the flow of the therapy session we based the control of events on a state diagram as shown in figure 4. At each state there are choices of next states that provide a natural progression in the flight sequence. Each state is of a fixed time duration that is long enough to provide a convincing animation but short enough that a transition to the next stage in the flight can happen within a relatively short time.

Those states that include links to themselves are static states, requiring no control from the therapist. For example, once the airplane has entered the nice flight state, it will remain in that state, circling around a simulated airport, until the therapist indicates that a new state should be entered. Those states without such self links are transitional states, and the simulator will remain in such states for only a limited amount of time. For example, once control has left the nice takeoff state, it will not enter that state again until the airplane has been flown through the flight, landing, and taxi states.

The simulator includes a simple menu that indicates the current state, as well as the list of transitions possible from the state. In static states, the link returning to the state is always followed, unless the therapist indicates that another link should be followed. In transitional states, control is very limited: the next state is predetermined. The therapist may end the simulation at any time.

In the engines off state, the airplane is parked at the airport terminal, without any sound effects. In the engines on state, the airplane is still parked at the terminal, but the engines sound effect is playing (and will continue playing until the engines off state is entered). In the taxi state, the preflight sound effect is played, and the airplane begins taxiing around the airport terminal. The nice takeoff state is a transitional state, in which the airplane moves from the taxi loop to the runway, plays the takeoff sound effect, and at the same time takes off, moving to the nice flight state. The rough takeoff state is similar to the nice takeoff state, but includes the thunder sound effect, flight turbulence, and a darkened sky. It moves to the rough flight state, where the thunder sound effect, turbulence and darkened sky continue until the nice flight, nice landing, or buzzing flight states are entered. In the nice flight state, the airplane circles around the airport in a calm manner.

The rough flight state is similar, but includes thunder, turbulence, and a darkened sky. The buzzing flight states makes close, low altitude passes over the airport, with accompanying engine (takeoff) sounds. The nice landing state is a transitional state, which moves from the nice flight state to the taxiing state by landing the airplane with accompanying prelanding and tire sound effects. The rough landing is a similar transitional state, which moves from the rough flight state to the taxiing state like the nice landing state, but which also includes thunder, turbulence and a darkened sky.

### Sound Effects

There are very specific sounds that are associated with flying on an airplane. In the design phase of the virtual airplane, we theorized that coordinating appropriate sounds with corresponding visual cues would be crucial to the sense of presence experienced by the user. Sounds were obtained from commercially available sound sample compact disc libraries, transferred to AIFF format on hard disk. These sounds were then edited for length and content. Our base software, the Simple Virtual Environment (SVE) Toolkit [13] includes a sound package which allows the programmer to play these sounds on either a local or a remote machine with a few simple function calls. The ability to play sounds on a remote machine both allowed us to distribute the computational load and is particularly useful if the local machine either lacks audio output or is physically located in a different room.

Sound effects were organized as discrete recordings that were either played once in coordination with events with a specified time interval or looped for continuous play. Specified time interval sound effects included the preflight and prelanding briefings, takeoff, and landing. Looped sound effects were the engine sounds and the storm sounds used to simulate inclement weather. Sounds were coordinated with visual effects. For example, when the storm sounds were used the sky outside the plane turned dark and the plane visually shook. Sounds could also be played simultaneously. For example, the engine sounds while landing were played simultaneously with the sounds of the tires touching down on the runway.

### Case Study

As an initial step in evaluating the design of the virtual airplane and its usefulness for fear of flying exposure therapy, we conducted a case study with one subject who suffered from fear of flying.

### Measures

We utilized a number of measures and techniques to assess the subject's attitudes and fear before, during, and after the therapy program. Every five minutes during therapy sessions the subject was asked to provide a subjective rating of her anxiety on a 0 (= completely calm) to 100 (panic level anxiety) scale. This Subjective Units of Discomfort (SUDs) rating system was also used by the subject to record her anxiety on an actual flight taken at the end of the therapy. Every therapy session was video-taped for review and analysis. Other measures used included:

The *Questionnaire on Attitudes toward Flying* (QAF) [12] assesses history of fear of flying, previous treatment, and attitudes toward flying. It includes a 36-item questionnaire rating the level of fear on a 0-11 scale in different flying situations. The possible range of scores on the QAF is 0-360. Test-retest reliability has been measured as 92% and split-half reliability as 99%.

The *Fear of Flying Inventory* (FFI) [18] is a 33-item scale measuring intensity of fear of flying. Items are rated on a 0-8 scale. Test-retest reliability has been measured as 92% and it has been sensitive to changes in behavior as a result of treatment.

The *Self-Survey of Stress Responses* (SSR) [6] is a 38-item measure tapping fearful responses during flying. Items are rated on a 0-5 scale and are divided into Autonomic (e.g., "My heart beats very fast"), Muscle ("My head aches.") or Central Process ("I bite my nails.") scores. The maximum score is 70 in each subscale.

The *Airplane Scenarios Self Ratings Sheet* (ASSRS) was developed for this project and lists eight scenarios: sitting on a plane–engines off, sitting on a plane–engines on, taxiing, takeoff, smooth flight, turbulent flight, thunderstorm and turbulent flight, and landing. Each scenario is rated on a 0 (no anxiety) to 100 (panic-level anxiety) scale.

The *Clinical Global Improvement (CGI) Scale* [9] is a global measure of change in severity of symptoms. The scale is bipolar with 1 = very much improved, 7 = very much worse, and 4 = no change. CGI has been used extensively in clinical trials for a variety of psychiatric patients.

The *Flight Self-Monitoring Sheet* was developed for use in this project to record subjective anxiety during the post-treatment flight. The subject was provided with a self-monitoring sheet to assess her Subjective Units of Discomfort (SUDs) Ratings at various points prior to boarding the plane, on board, in flight, and afterwards.

The *Therapist Form* was constructed for this project to record virtual reality situations presented to the client during therapy sessions, SUDs ratings during exposure, session-by-session homework, clinical global improvement, and subject self-rating of global improvement.

### Subject

The subject was a 42-year-old female with a debilitating fear and avoidance of flying, who met clinical criteria [1] for a specific phobia. She was working gainfully in a profession that did not require travel. She reported a prior history of regular but fearful flying for vacations, having grown increasingly fearful over the past five years, worrying about the plane crashing. During this period she used antihistamines or anxiolytic medication to allow her to fly. She ceased flying completely two years prior to seeking treatment. She sought treatment because she missed the family vacations and was entering into a new business venture that would require airplane travel.

### Procedure

The subject was referred to one of the authors (Rothbaum) for treatment of the fear of flying. She was administered the battery of questionnaires described above and seen as a paying outpatient for seven sessions during which time she was taught anxiety management techniques (AMT). At this point in the therapy, it was mutually decided that exposure would be beneficial. The subject was offered free treatment using virtual reality exposure (VRE) and accepted.

Immediately prior to the first VRE session, the subject was informed of the study procedures and the experimental nature of the treatment. At this time the same battery of questionnaires was completed and an informed consent form was signed. The first treatment session of VRE followed. Sessions were held twice weekly with each session lasting approximately 35-45 minutes. Six treatment sessions were conducted. During each therapy session the subject was allowed to progress at her own pace along a natural progression of scenarios: sitting on the plane with engines off, sitting on a plane with engines on, taxiing, smooth take-off, smooth flight, close pass over the airport similar to a missed landing, landing, rough take-off, turbulent flight and rough landing. The therapist viewed all virtual environments to which the subject was exposed on a monitor and thus was able to make appropriate comments and encourage continued exposure until anxiety decreased.

Immediately following the last VRE session, the subject completed the battery of questionnaires a third time. Two days following the last VRE session the subject completed a planned cross country flight with her family (husband and two children). During each of the two legs for the initial and return flights, she completed the Flight Self Monitoring Sheet. One month after completion of therapy, the battery of questionnaires was administered as a follow-up assessment.

## RESULTS

### Presence

One of our interests in this study was the degree in which the virtual airplane would cause the subject to experience a sense of presence of being on a real aircraft. Unfortunately, presence is a subjective sensation and no direct measures of a user's sense of presence that are operational, reliable and useful have been developed [10, 19]. We can however, record the subject's verbal and physical responses to the experience as indirect evidence of the user's sense of presence.

The subject experienced many of the classic anxiety symptoms that would be manifested by a person on a real airplane who was afraid of flying. During the six sessions, our subject experienced tightness in the chest, shortness of breath, butterflies in her stomach, and tenseness in her shoulders and stomach. In addition, the level of her anxiety paralleled the events in her virtual flight. For example, her anxiety went up when the virtual airplane flew through a storm and took off.

Another indirect measure of the subject's sense of presence is the usefulness of the virtual airplane for therapy. Discussions of emotional processing theory as applied to anxiety disorders [4, 5] purport that fear memories can be construed as structures that contain information regarding stimuli, responses, and meaning. Therapy is aimed at facilitating emotional processing. For this to occur, it has been proposed that the fear structure must be activated and modified. For virtual environments to be effective, they must activate the fear structure and elicit the fearful responses. Evoking a sense of presence on an airplane that is strong enough for the patient to become anxious is therefore essential to conducting successful exposure therapy.

### Clinical Results

The pre- and post-treatment scores are shown in Table 1. The subjects self-reported anxiety about flying was reduced following training in anxiety management techniques (AMT), decreased further following VR exposure to flying, and was maintained at follow-up.

**TABLE 1: TEST SCORES**

| Measure | | Before Therapy | After in-office anxiety management techniques | After VR Exposure | One-month post treatment |
|---|---|---|---|---|---|
| QAF | 36-items | 215 | 113 | 62 | 74 |
| FOFI | | 93 | 88 | 50 | 53 |
| SSSR: | A | 22 | 14 | 17 | 17 |
| | M | 13 | 9 | 7 | 7 |
| | C | 9 | 7 | 8 | 6 |
| ASSRS | Engine Off | 10 | 10 | 5 | 10 |
| | Engine On | 20 | 10 | 5 | 15 |
| | Taxiing | 30 | 20 | 5 | 15 |
| | Takeoff | 40 | 30 | 10 | 20 |
| | Smooth Flight | 50 | 30 | 10 | 15 |
| | Turbulent Flight | 80 | 50 | 15 | 25 |
| | Turbulent-Storm | 100 | 70 | 20 | 40 |
| | Landing | 40 | 20 | 10 | 15 |

Note: QAF= Questionnaire on Attitudes toward Flying; FFI= Fear of Flying Inventory; SSSR= Self-Survey of

**TABLE 2. SUDS RATINGS DURING POST-TREATMENT FLIGHT**

| | Flight Out | | Flight Back | |
|---|---|---|---|---|
| Situations | 1st Leg | 2nd Leg | 1st Leg | 2nd Leg |
| Pre-board | 15 | 20 | 10 | 15 |
| Pre-takeoff | 20 | 15 | 15 | 20 |
| 5 min. after takeoff | 25 | 10 | 30 | 25 |
| 5 min. before landing | 20 | 15 | 20 | 15 |
| Landing | 15 | 10 | 10 | 10 |
| Post-landing | 10 | 5 | 5 | 5 |

Note: SUDs = Subject Units of Discomfort scale, 0 = no anxiety, 100 = maximum anxiety.

The subject's self-reported fear decreased from "8" at pre-treatment to "4" at post-treatment on a 0-10 scale of fear of flying on the QAF, and the therapist and subject rated her as "much improved" (2 on the CGI). In addition she was able to complete a round-trip cross country flight with minimal anxiety immediately following treatment. Table 2 contains SUDs ratings during the four legs of this post-treatment flight. During these flights, her highest SUDs rating was 30. Some of her comments written on the flight self-monitoring sheets include:

"Still have blips of anxiety on turbulence, but less intense this leg."

"About ten minutes of strong turbulence but I stayed pretty calm... Somehow it was comforting at points to remember and picture the VR."

"Much better than in last few years but not yet perfect."
Virtual reality exposure treatment was successful in reducing this subject's fear of flying. Although the sole contribution of VR exposure to her improvement is not possible to determine given the inclusion of AMT techniques, the contribution of VR exposure to the overall outcome is considered significant for many reasons. First, her self-reported anxiety on all measures decreased further following VR exposure. Second, she was able to complete a long flight, one that she had avoided for the past two years. Third, the usual treatments for fear of flying include a combination of AMT and exposure. Had VR exposure not been available, this therapy would have included exposure to actual airplanes, possibly with the therapist flying with the patient before she flew with her family. Instead the treatment goals were accomplished using VR exposure instead of *in vivo* exposure, which is the significance of this case report to clinical psychology. A more detailed analysis of the clinical aspects of this study are available in [17].

## Implications

There are a great many things that we can not do well in VR. Visually, virtual environments are still cartoon worlds and are not very realistic. The combination of current tracker technology and the graphics pipeline guarantee a lag between head movement and response of the visual image. The spaces that we can track are small and tracker accuracy is poor. Haptic cues in virtual environments are usually nonexistent or very limited.

As a result, examples of VR applications that offer sufficient value beyond that available from less exotic technology are still rare.

To create successful applications with today's VR technology we must begin by asking: *what are we good at?* The obvious answer is that we are good at giving a person the illusion of actually inhabiting a computer-generated space. The defining factor that distinguishes a virtual reality system from an interactive computer graphics or multimedia system is the immersion of the user in a computer-generated environment and the resulting sense of presence she experiences.

To build successful applications, especially with current technology, we must next ask: *what are the applications whose success is dependent on the user's sense of presence in a computer-generated environment?* Virtual reality exposure therapy is one such application. This study and our previous studies [11, 15, 16] have provided clinical evidence that use of virtual environments for exposure therapy is useful and has advantages over competing therapy modalities. For exposure therapy to be effective, the subject must be exposed to environments that activate his fear. We have shown that current VR technology combined with carefully designed environments can successfully be used to induce anxiety in a patient, that repeated exposure leads to anxiety reduction, and that therapy conducted within virtual environments transfers to real, physical situations. This case study has also provided evidence for the usefulness of virtual environments as a training tool to practice anxiety management skills. In addition to being effective, VR therapy also provides advantages in terms of therapist's time, control of the therapy environment, risk management, patient confidentiality and client willingness to undergo therapy over conventional treatment modalities.

## Acknowledgments

Eric Brittain, Mave Houston and Marsha Brackner created the original model for the virtual airplane. Eric also did preliminary work on finding and mixing the sound effects. Major James Williford, Division Psychiatrist for the 101st Airborne, provided many helpful discussions on current fear of flying therapy approaches. This work was partially supported by a National Science Foundation Research Experiences for Undergraduates Site Grant and by a grant from Hewlett-Packard.